\documentclass[aps,prl,preprint,superscriptaddress]{revtex4-1}

\usepackage{natbib}
\usepackage{amsmath}
\usepackage[toc,page]{appendix}
\usepackage{bm}
\usepackage{ amssymb }
\usepackage{graphicx}
\usepackage {epstopdf}
\usepackage{csquotes}

\begin{document}


\title{High temperature resistivity measured at $\nu=\frac{5}{2}$ as a predictor of 2DEG quality in the N=1 Landau level}


\author{Q. Qian}
\affiliation{Department of Physics and Astronomy, Purdue University}
\affiliation{Station Q Purdue, Purdue University}

\author{J. Nakamura}
\affiliation{Department of Physics and Astronomy, Purdue University}
\affiliation{Station Q Purdue, Purdue University}

\author{S. Fallahi}
\affiliation{Department of Physics and Astronomy, Purdue University}
\affiliation{Station Q Purdue, Purdue University}
\affiliation{Birck Nanotechnology Center, Purdue University}

\author{G. C. Gardner}
\affiliation{Station Q Purdue, Purdue University}
\affiliation{Birck Nanotechnology Center, Purdue University}
\affiliation{School of Materials Engineering, Purdue University}
\author{J. D. Watson}
\affiliation{Department of Physics and Astronomy, Purdue University}
\affiliation{Birck Nanotechnology Center, Purdue University}




\author{M. J. Manfra}
\email[]{mmanfra@purdue.edu}
\affiliation{Department of Physics and Astronomy, Purdue University}
\affiliation{Station Q Purdue, Purdue University}
\affiliation{Birck Nanotechnology Center, Purdue University}
\affiliation{School of Materials Engineering, Purdue University}
\affiliation{School of Electrical and Computer Engineering, Purdue University}

\date{\today}

\begin{abstract}
We report a high temperature (T = 0.3K) indicator of the excitation gap $\Delta_{5/2}$ at the filling factor $ \nu=5/2$ fractional quantum Hall state in ultra-high quality AlGaAs/GaAs two-dimensional electron gases.  As the lack of correlation between mobility $\mu$ and $\Delta_{5/2}$ has been well established in previous experiments,  we define, analyze and discuss the utility of a different metric $\rho_{5/2}$, the resistivity at $\nu=5/2$, as a high temperature predictor of $\Delta_{5/2}$. This high-field resistivity reflects the scattering rate of composite fermions. Good correlation between $\rho_{5/2}$ and $\Delta_{5/2}$ is observed in both a density tunable device and in a series of identically structured wafers with similar density but vastly different mobility. This correlation can be explained by the fact that both $\rho_{5/2}$ and $\Delta_{5/2}$ are sensitive to long-range disorder from remote impurities, while $\mu$ is sensitive primarily to disorder localized near the quantum well. 
\end{abstract}
\maketitle
The two-dimensional electron gas (2DEG) in GaAs/AlGaAs heterostructures remains the preeminent semiconductor platform for the study of strong electron-electron correlations in reduced dimensions.  As the design of GaAs/AlGaAs 2DEG structures becomes more sophisticated and ultra-low temperature experiments become more complicated, the question of how best to make a preliminary assessment of sample quality becomes increasingly important \cite{DasSarma2014}.  This is especially true for heterostructures designed to explore the most fragile fractional quantum Hall states (FQHSs) in the N=1 Landau level (LL), a regime in which many distinct phases are separated by small intervals in filling factor $\nu=hn/eB$ (h is Planck's constant, n is 2DEG density, e is the charge of the electron and B is magnetic field) that must be examined at temperatures below 50mK. For example, transport signatures of the putative non-Abelian $\nu=5/2$ and $\nu=12/5$ FQHS and reentrant integer quantum Hall effect (RIQHE) states are strongest at temperatures T$\leq$20mK. Traditionally, zero-magnetic-field mobility measured at much higher temperatures (0.3K$\leq$T$\leq$2K) has been used as a primary metric of 2DEG quality, but a large body of experimental and theoretical evidence has now shown that mobility does not necessarily encode all information needed to predict high-field behavior in the fractional quantum Hall regime\cite{Neubler2010, Gamez, Manfrareview, DasSarma2014, Pan2011}. 
Evidently additional methods must be employed to predict behavior at lower temperatures and high magnetic field in the highest quality samples \cite{Manfrareview}. In the context of the present work, quality is quantified by the size of the energy gap of the FQHS at $\nu=5/2$, $\Delta_{5/2}$. We note that samples with high $\Delta_{5/2}$ generally show high-quality transport throughout the N=1 LL.
\begin{figure}[t]
\def\ffile{Fig1.pdf}
\centering
\includegraphics[width=0.6\linewidth]{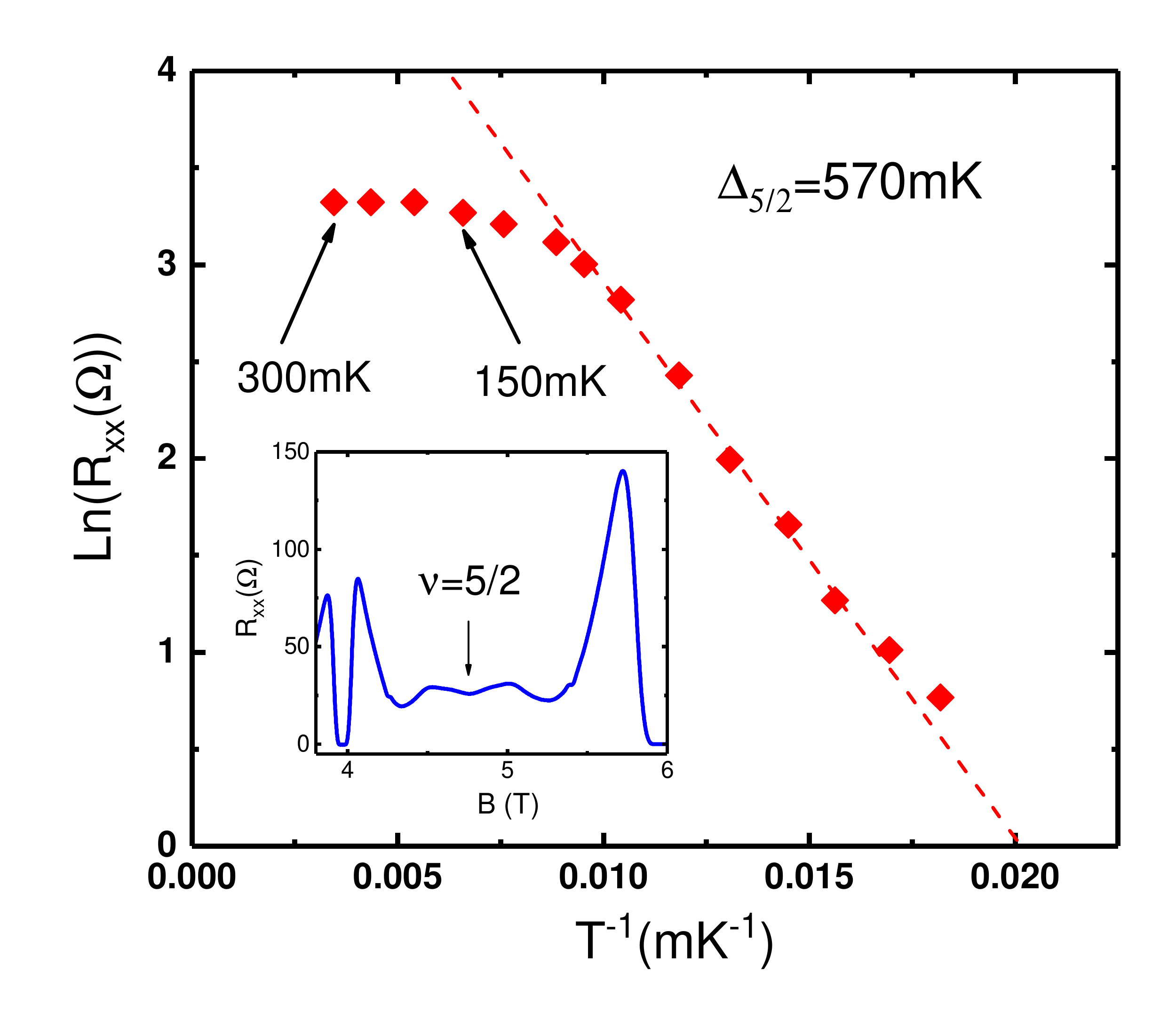}
\caption{\label{Arrhenius} (color online).  Arrhenius plot at $\nu=5/2$ where the gap is measured to be 570mK. Magnetoresistance at T=0.3K in a larger field range is shown in the inset.}
\end{figure}

Composite fermion (CF) theory replaces a system of highly interacting electrons with a system of weakly-interacting composite fermions by vortex attachment \cite{Halperin1993, Jain1, Jain2} and explains the physics around filling factor $\nu=1/2$ in the N=0 LL. Extending this theory to other half-fillings in higher LLs, we expect that at T=0.3K, the state at $\nu=5/2$ is  also described by composite fermions experiencing zero effective magnetic field $B_{eff}=0$. Indeed, Willett and collaborators \cite{BoB} demonstrated the existence of a Fermi surface at $\nu=5/2$ at T=0.3K using surface acoustic wave techniques. 
The CF model has also been successfully used to analyze energy gaps of odd denominator FQHSs  around $\nu=5/2$ by using the CF cyclotron energy \cite{Kumar, Samkharadze}. In the CF picture, the fractional quantum Hall state that emerges at T$\leq$100mK is viewed as a condensation of CFs driven by a BCS-like p-wave pairing instability \cite{Moore1991, JainBook}.  In this work, we assume  that at T=0.3K a Fermi sea of CFs forms at $\nu=5/2$, and measure the resistivity of this metallic phase,  $\rho_{5/2}$,  analogously to the zero field resistivity. We investigate this high field resistivity as a high temperature (T = 0.3K) indicator of the strength of the $\nu=5/2$ FQHS at low temperatures.

The longitudinal resistance as a function of magnetic field B measured at T = 0.3K in the vicinity of $\nu=5/2$ is plotted in the inset of Fig. \ref{Arrhenius}. A resistance minimum is observed at $\nu=5/2$. It is noteworthy that the positive magnetoresistance near $\nu=5/2$ resembles the transport behavior near $\nu=3/2$ and $\nu=1/2$ \cite{Du1993, Rokhinson1995, Rokhinson1997}, but contrasts with the negative magnetoresistance often observed near zero field. The temperature dependence of resistance at $\nu=5/2$ is shown in Fig. \ref{Arrhenius}. In this exemplary Arrhenius plot, $R_{xx}$ at $\nu=5/2$ shows activated behavior below 100mK: it increases as temperature increases following $R_{xx}\varpropto e^{-\frac{\Delta}{2 k_{B}T}}$. A linear fit through the activation region yields an energy gap $\Delta_{5/2}$=570mK. However, $R_{xx}$ at $\nu=5/2$ starts to saturate when temperature exceeds 150mK, and at T=300mK it is insensitive to temperature. The very weak temperature dependence observed around T=0.3K is an important attribute; it indicates that a description based on a gapped FQHS with a dilute population of thermally activated FQHS quasiparticles is no longer justified as it is at significantly lower temperatures.  It is appealing to consider this change a temperature-driven transition to a CF Fermi sea. As a purely practical matter, the temperature insensitivity of $\rho_{5/2}$ suggests the utility of our method of characterization in much the same way that B=0 mobility is insensitive to temperature below T$\sim$1K in ultra-high quality samples. 
The resistivity measured at $\nu=1/2$ is equal to the CF resistivity. It can be shown that the resistivity measured at $\nu=5/2$ is equivalent to CF resistivity up to a numerical scale factor \cite{Rokhinson1995,Halperin1993}. 

 
We have studied two types of samples. The first sample is a density tunable device: an $\textit{in situ}$ back-gated 2DEG. The 2DEG is grown by molecular beam epitaxy (MBE) and resides in a 30nm GaAs quantum well bounded by Al$_{0.24}$Ga$_{0.76}$As barriers with Si $\delta$-doping in a narrow GaAs layer flanked by pure AlAs layers placed 66nm above the principal 30nm GaAs quantum well. This design has been shown to yield the largest gap energy for the $\nu=5/2$ FQHS \cite{Manfrareview, Deng1, Deng2, Josh}. It is believed \cite{Neubler2010} that low conductivity electrons confined in the X-band of the AlAs barriers surrounding the narrow GaAs doping wells screen potential fluctuations caused by remote donor impurities, promoting expression of large gap FQHS in the N=1 LL. This particular sample displays the largest excitation gap for the $\nu=5/2$ FQHS reported in the literature, attesting to its high quality \cite{Waston}. The $\textit{in situ}$ gate consists of an $n^{+}$ GaAs layer situated 850 nm below the bottom interface of the quantum well. There is an 830nm  GaAs/AlAs superlattice between the 2DEG and back gate to minimize leakage current. The device is a 1mm by 1mm lithographically-defined Van der Pauw (VdP) square with eight Ni$\setminus$Ge$\setminus$Au$\setminus$Ni stack contacts diffused along the sample edges;  processing details have been described in reference \cite{Waston}. 

In a second set of experiments we examine a series of samples each sharing the same heterostructure design: a 30nm GaAs quantum well with 2DEG density $ n=3.0\times$ $ 10^{11}/cm^{2}$. The quantum well is flanked by symmetric Si $\delta$-doping in GaAs doping wells as described previously \cite{Manfrareview}. These samples are grown in a single MBE growth campaign; sample mobility improves as the unintentional impurities emanating from the source material decrease with increasing growth number. Each specimen consists of a 4mm by 4mm VdP square with 8 diffused indium contacts on the edges. 
We perform standard low frequency (13-85 Hz) lock-in measurements. The excitation currents for resistivity measurement and gap measurement are 10nA and 2nA,  respectively. We use the Van der Pauw method for measuring resistivites; the quoted resistivity is the usual average of four distinct resistance measurements. The samples are homogeneous such that the resistances measured along different crystallographic directions are similar, both at $\nu=5/2$ and zero field.

\begin{figure}[t]
\def\ffile{Fig2.pdf}
\centering
\includegraphics[width=0.6\linewidth]{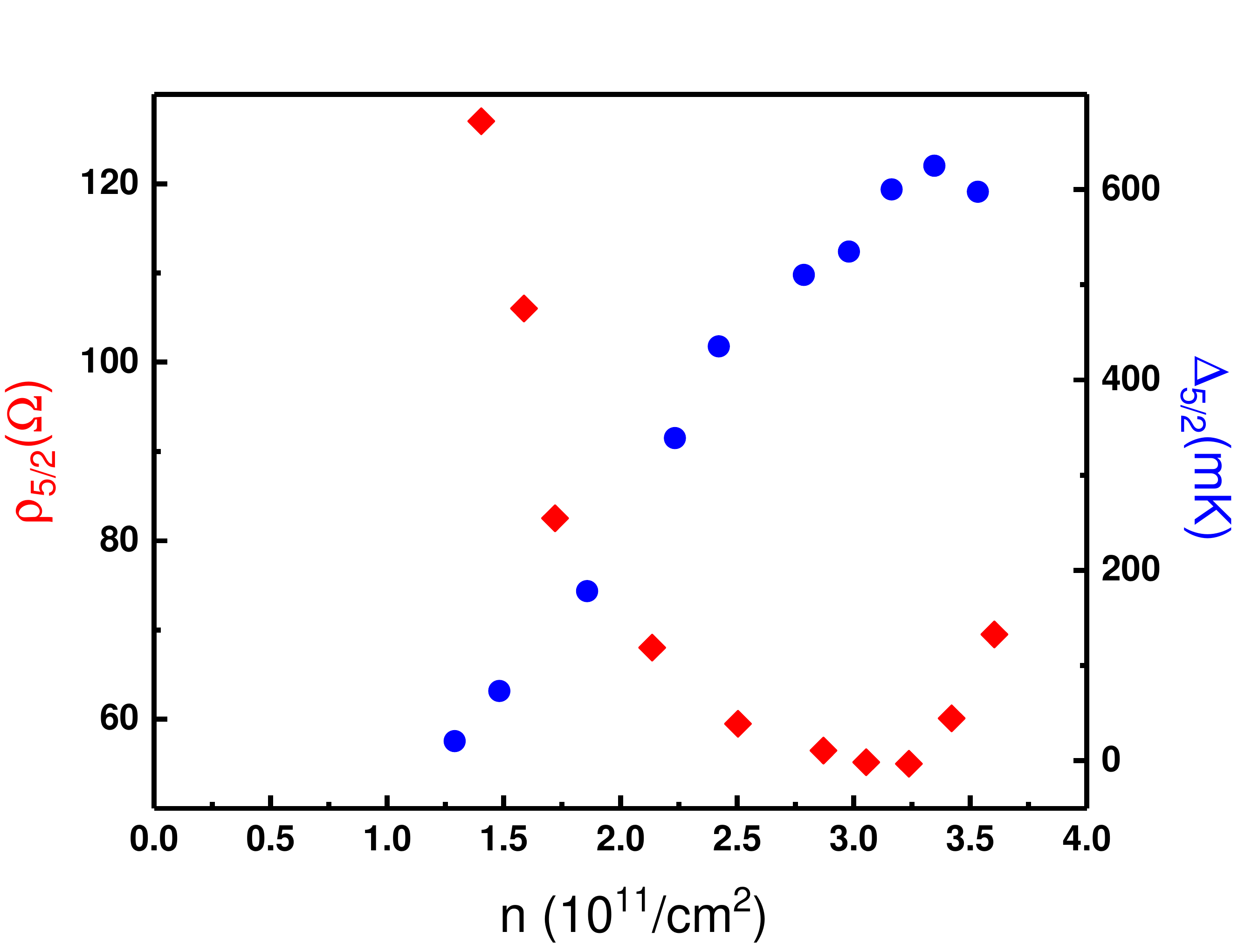}
\caption{\label{BG_5halfR} (color online). $\nu=5/2$ resistivity $\rho_{5/2}$ measured at T = 0.3K (Red) and $\Delta_{5/2}$ (Blue) as a function of 2DEG density $n$ for the  $\textit{in situ}$ back gated sample.}
\end{figure}
In Fig. \ref{BG_5halfR}(a) we present the dependence of $\rho_{5/2}$ (left axis) and $\Delta_{5/2}$ (right axis) for various densities of the $\textit{in situ}$ back gated GaAs 2DEG. The typical uncertainty for gap and resistivity measurement is $\pm5 \%$. As $n$ increases, $\Delta_{5/2}$ increases and $\rho_{5/2}$ decreases. A clear correlation between $\rho_{5/2}$ and $\Delta_{5/2}$ is observed in this density tunable device.  

In the density-tunable device, we expect that as density increases the resistivity $\rho_{5/2}$ should decrease due to the increasing carrier concentration, and the energy gap $\Delta_{5/2}$ should increase due to the increase of the Coulomb energy scale. It is also possible that the scattering rate may change with changing density \cite{Halperin1993, DasSarma2014}, so the change of $\rho_{5/2}$ and $\Delta_{5/2}$ with changing density likely reflects both their explicit density dependence and the effects of scattering. The strong correlation we observe between $\rho_{5/2}$ and $\Delta_{5/2}$ in this device indicates that $\rho_{5/2}$ captures both the density dependence and the effects of scattering on the energy gap.

We note that at high densities, $\rho_{5/2}$ begins to increase and $\Delta_{5/2}$ plateaus and decreases slightly. This has been explained by occupation of the first excited subband of the quantum well \cite{Waston}, and may be responsible for the slight mismatch between the minimum of $\rho_{5/2}$ and the peak of $\Delta_{5/2}$; evidently the correlation between $\rho_{5/2}$ and $\Delta_{5/2}$ is not as strong when there is parallel conduction through an excited subband.

\begin{figure}[t]
\def\ffile{Fig3.pdf}
\centering
\includegraphics[width=0.65\linewidth]{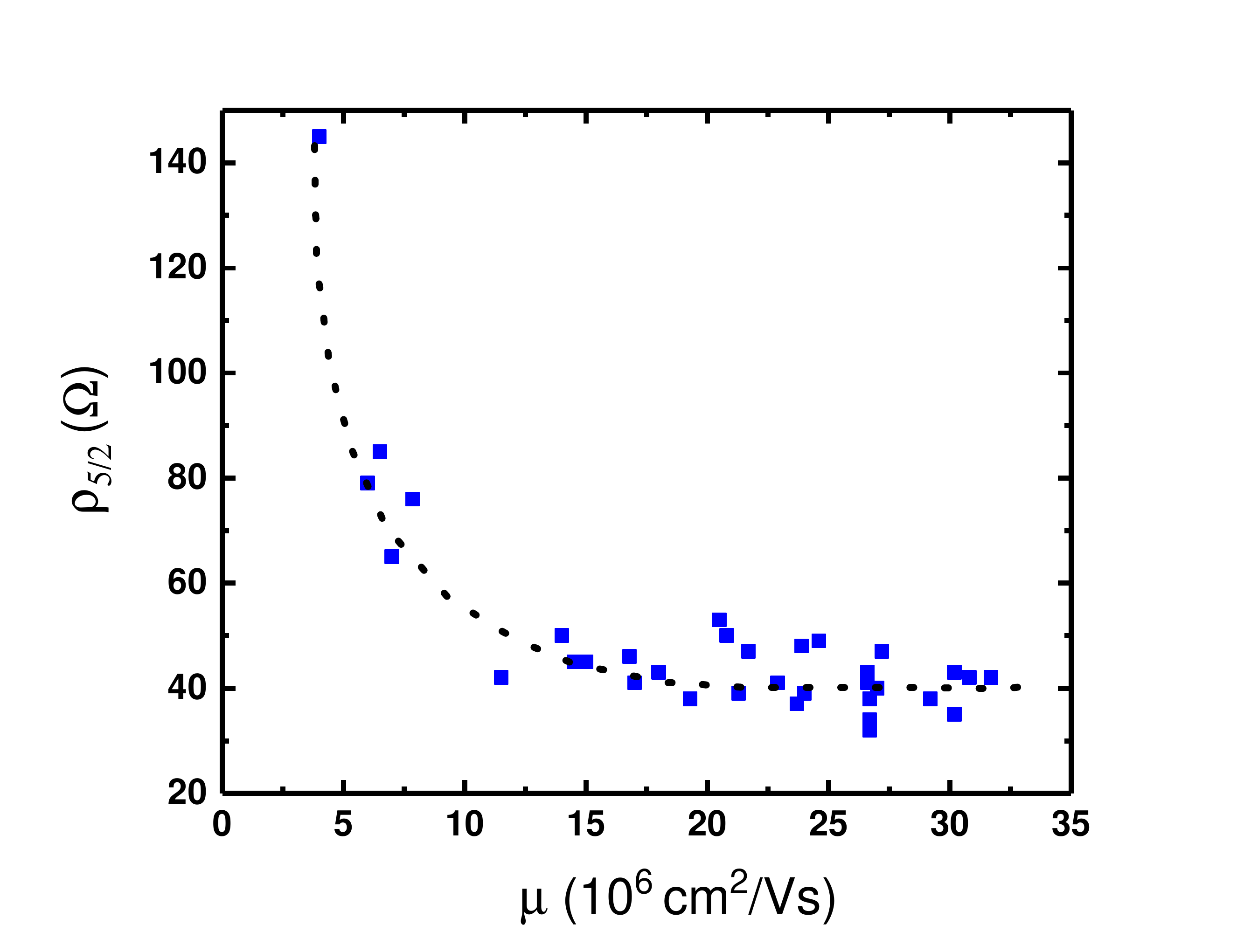}
\caption{\label{5half_resistivity_vs_mu}  (color online). $\nu=5/2$ resistivity $\rho_{5/2}$ measured at T=0.3K vs. mobility for samples grown during a single MBE growth campaign. All samples have same heterostructure design: a symmetrically doped GaAs quantum well with density $n=3.0\times$ $ 10^{11}/cm^{2}$. The dashed line is guide to the eye.}
\end{figure}

We then measured $\rho_{5/2}$ for the series of wafers with the identical heterostructure design and fixed 2DEG density.
The relationship between $\rho_{5/2}$ and $\mu$ is illustrated in Fig. \ref{5half_resistivity_vs_mu}: $\rho_{5/2}$ initially drops monotonically as mobility increases, but it saturates at $\mu\sim 15\times10^6cm^2/Vs $ even though mobility keeps increasing over the course of the MBE growth campaign. In this high mobility range $\rho_{5/2}$ and $\mu$ appear to have no relationship to one another; samples with the same $\rho_{5/2}$ may have vastly different $\mu$.
\begin{figure}[t]
\def\ffile{Fig4.pdf}
\centering
\includegraphics[width=0.8\linewidth]{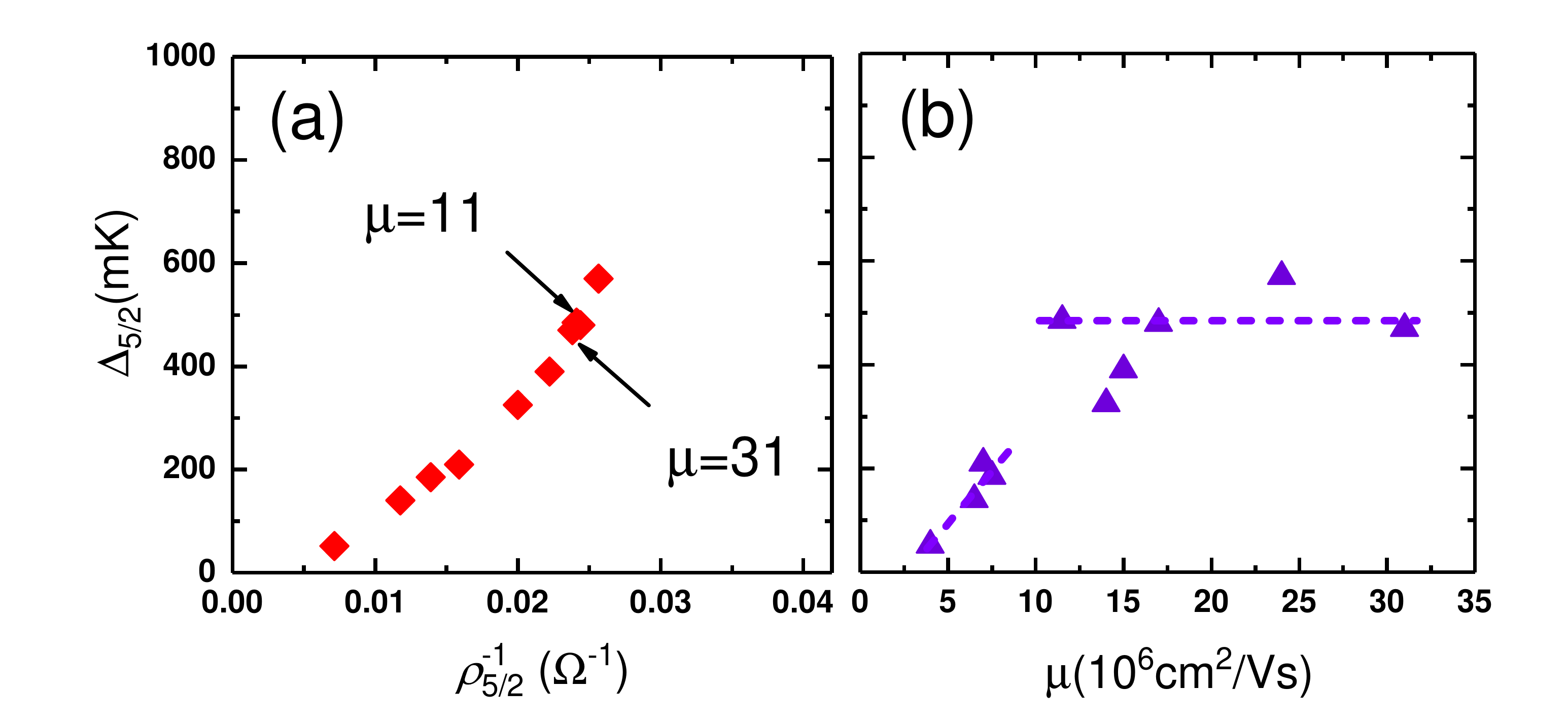}
\caption{\label{5halfgap} (color online). The $\nu=5/2$ energy gap $\Delta_{5/2}$ vs. 1/$\rho_{5/2}$ (a) and mobility $\mu$ (b) for samples chosen from Fig. \ref{5half_resistivity_vs_mu}. The arrows in (a) indicate two samples with the same $\rho_{5/2}$ and $\Delta_{5/2}$ but vastly different $\mu$, and the units for the listed $\mu$ are $10^6$ $cm^2/Vs$. The dashed lines in (b) are guides to the eye. }
\end{figure}
A few samples with various combinations of $\rho_{5/2}$ and $\mu$ are chosen from the sample set in Fig. \ref{5half_resistivity_vs_mu} to measure energy gaps ($\Delta_{5/2}$) of the fractional quantum Hall state that forms at $\nu=5/2$ at lower temperatures. Fig. \ref{5halfgap} (a) displays the relation between $\Delta_{5/2}$ and $1/\rho_{5/2}$ for those samples: $\Delta_{5/2}$ increases monotonically as $1/\rho_{5/2}$ increases. For the largest $1/\rho_{5/2}$, where $\rho_{5/2}$ is $39 \Omega$, $\Delta_{5/2}$ reaches 570mK, among the largest gaps at this density ever reported in literature \cite{Deng1, Waston,Kumar}. Here, in this comparison of different samples with the same density but different levels of disorder,  we also observe a strong correlation between $\Delta_{5/2}$ and $\rho_{5/2}$, indicating that $\rho_{5/2}$ is sensitive to the scattering mechanisms that limit $\Delta_{5/2}$. 
We also plot $\Delta_{5/2}$ vs. $\mu$ for these samples. As it is shown in Fig. \ref{5halfgap} (b),  in the low mobility range where $\mu < 10 \times 10^6 cm^2/ Vs$, $\Delta_{5/2}$ increases as $\mu$ increases. However, no correlation exists in the high mobility range where $\mu > 10 \times 10^6 cm^2/ Vs$. 

We briefly discuss why the CF resistivity measured by $\rho_{5/2}$ may contain different information than the zero-field mobility and show a stronger correlation with the low-temperature $\nu=5/2$ FQHS. Zero-field resistivity is dominated by large-angle scattering [1]. Because of this, the zero-magnetic-field mobility is primarily limited by Coulomb scattering from impurities located directly in the quantum well \cite{DasSarma2008, DasSarma2014, Deng1, Gardner2016}, while remote impurities primarily cause small-angle scattering which has significantly less impact on mobility. Composite fermions at half-filling are also scattered by impurities; however, remote charged impurities also induce spatial variations in the effective magnetic field $B_{eff}$ due to variations in the 2DEG density \cite{Halperin1993, Zhang1992}. These effective magnetic field variations cause increased large-angle scattering of CFs, resulting in an enhanced contribution by remote impurities to the CF resistivity \cite{Halperin1993, Rokhinson1995, Mirlin1995}. Because of this, we expect that $\rho_{5/2}$ is more sensitive to remote impurities than the zero-field mobility and thus contains different information about the distribution of impurities in the system. This increase in large-angle scattering has been observed experimentally at $\nu=1/2$ \cite{Du1993}.  However, the connection between CF resistivity at $\nu=5/2$ and FQHS gap strength has not been previously studied. Our data suggests that variations of $B_{eff}$ from remote impurities dominate the measured $\rho_{5/2}$.

Other experiments \cite{Rokhinson1995, Rokhinson1997, Kang1995} and theory \cite{Khveshchenko1996, Mirlin1997} at $\nu=1/2$ have shown that short-range CF-CF interactions via gauge field fluctuations result in an additional correction to the CF scattering rate and resistivity which is not present at zero field. This may be another reason that $\rho_{5/2}$ provides information about the disorder potential from impurities that $\mu$ does not. 

Next, we address the sensitivity of the strength the $\nu=5/2$ FQHS to disorder. It has been shown through computational methods that the size of quasiparticles and quasiholes in the $\nu=5/2$ state is unusually large, on the order of at least 12 times the magnetic length \cite{Neubler2010}; this large size is expected to make $\Delta_{5/2}$ sensitive to long-range disorder from remote charged impurities \cite{Neubler2010, Pan2011}. Experiments studying the effects of remote doping schemes have confirmed that remote impurities from ionized donors do indeed have a large impact on $\Delta_{5/2}$, but minimal effect on $\mu$ \cite{Gamez, Umansky2009}. A particularly illuminating experiment is detailed in Ref. \cite{Deng1}: it was found that intentionally placing short-range disorder directly in the quantum well drastically reduced mobility but had a comparatively small effect on $\Delta_{5/2}$, confirming that $\mu$ is limited by short-range disorder while $\Delta_{5/2}$ is more sensitive to long-range disorder from remote impurities. The fact that both $\Delta_{5/2}$ and $\rho_{5/2}$ are sensitive to long-range disorder from remote impurities explains the strong correlation we observe between the two quantities and the lack of correlation with $\mu$.

Additionally, if the $\nu=5/2$ FQHS is considered to be a p-wave Cooper-paired state of composite fermions as in the Moore-Read Pfaffian state \cite{Moore1991, JainBook}, then it is natural to compare our results to what is observed in p-wave superconductors. In p-wave superconductors, as the normal-state resistivity increases due to impurity scattering, the superconducting transition temperature $T_c$ is expected to decrease \cite{Abrikosov1960, Millis1988, Levin1993, Mackenzie2003}, and strong suppression of $T_c$ with increasing normal-state resistivity has been observed experimentally in the putative p-wave superconductor Sr$_2$RuO$_4$ \cite{Mackenzie1998, Mackenzie2003}. Therefore, the direct correlation we observe between the normal-state CF resistivity at T=0.3K and the low-temperature $\nu=5/2$ FQHS energy gap is in qualitative agreement with the strong dependence of $T_c$ on normal-state resistivity in p-wave superconductors.        

We mention that the quantum scattering time $\tau_q$ measured by low-field Shubnikov-de Hass oscillations \cite{Coleridge1989, Coleridge1991} is also expected to be sensitive to long range disorder, and thus might be expected to be a predictor of $\Delta_{5/2}$ \cite{DasSarma2014}. However, a previous experiment in a density-tunable device \cite{Neubler2010} showed no correlation between $\tau_q$ and the strength of the $\nu=5/2$ FQHS. We also measured $\tau_{q}$ in our back-gated device and found no correlation with $\Delta_{5/2}$ (data not shown here); a detailed analysis of quantum scattering time in our samples is presented in a forthcoming publication. 

In conclusion, we observe a strong correlation between $\rho_{5/2}$ and $\Delta_{5/2}$ in both a density-tunable device and in a series of samples with fixed 2DEG density. Therefore, we propose the use of
$\rho_{5/2}$ measured at T = 0.3K as a metric to predict the strength of $\nu=5/2$ FQHS at low temperatures. The fact that we observe this correlation both when the electron density is varied (in the back-gated device) and when the level of disorder is varied (in the series of fixed-density samples) makes our method a robust tool for predicting $\Delta_{5/2}$. A possible physical origin for the correlation is that $\rho_{5/2}$ is sensitive to large-angle scattering by remote impurities due to the variation of the $B_{eff}$ and to short-range CF-CF interactions, neither of which are reflected in the zero-field mobility. 

This work was supported by the Department of Energy, Office of Basic Energy Sciences, under Award number DE-SC0006671.  Additional support for sample growth from the W. M. Keck Foundation and Microsoft Station Q is gratefully acknowledged.  We thank S. H. Simon for important comments on this work.

\end{document}